# Three innovations of next-generation antibiotics: evolvability, specificity, and non-immunogenicity

Hyunjin Shim[1,*]

## Author Information

### Affiliations

[1]**Center for Biosystems and Biotech Data Science, Ghent University Global Campus, Incheon 21985, South Korea**

[*]Corresponding Author: Hyunjin Shim (jinenstar@gmail.com)

### Orcid links
Hyunjin Shim orcid=0000-0002-7052-0971

# Abstract

Antimicrobial resistance is a silent pandemic that is being exacerbated by the uncontrolled use of antibiotics. Since the discovery of penicillin, we have been largely dependent on microbe-derived small molecules to treat bacterial infections. For the last 100 years, this remedy has been extremely successful and the recovery from bacterial infections has been taken for granted through the extensive and profuse use of these antibiotics. However, the golden era of antibiotics is coming to an end as the emergence and spread of antimicrobial resistance against these antibacterial compounds is outpacing the discovery and development of new antibiotics. The current antibiotic market suffers from various shortcomings, including the absence of profitability and investment. However, the most important underlying issue of these traditional antibiotics arises from the inherent properties of these small molecules being mostly broad-spectrum and non-programmable. While bacteria adapt rapidly to evade and counteract the mechanisms of these small molecule inhibitors, the antibiotic market is still focusing mainly on these traditional antibiotics to tackle multidrug-resistant bacteria. As the scientific knowledge of microbes and microbial communities progresses, the scientific community is starting to explore entirely novel approaches to tackling the problem of antimicrobial resistance. One of the most prominent approaches is to develop next-generation antibiotics, which we define as alternative antimicrobial agents to small molecules that inhibit growth or kill bacteria. In this review, we discuss three innovations of next-generation antibiotics as compared to traditional antibiotics as specificity, evolvability, and non-immunogenicity. We present a number of potential antimicrobial agents in research that possess some innovative properties of being specific, evolvable, and non-immunogenic. These examples include bacteriophage-based therapy, CRISPR-Cas-based antimicrobials, and microbiome-derived antimicrobial agents. These alternative antimicrobial agents possess innovative properties that may overcome the inherent shortcomings of traditional antibiotics, and some of these next-generation antibiotics are not merely far-fetched ideas but are currently in clinical development. We further discuss some related issues and challenges such as infection diagnostics and regulatory frameworks that still need to be overcome to bring these next-generation antibiotics to the antibiotic market as viable products in combating antimicrobial resistance using a diversified set of strategies.

# 1. Introduction

A growing number of bacterial infections such as salmonellosis, tuberculosis, pneumonia, and gonorrhea are becoming resistant to antibiotics. The World Health Organization (WHO) recently declared the spread of antimicrobial resistance (AMR) as one of the top 10 threats to global health and development, which shows that the problem of multidrug-resistant bacteria is having a negative impact on various aspects of society. Followingly, the WHO published a priority list of pathogens that urgently require new antibiotics, which includes the most critical group that is prevalent in healthcare facilities, like *Acinetobacter*, *Pseudomonas*, and Enterobacteriaceae species (Table 1). In 2019, at least 1.2 million people died worldwide from multidrug-resistant bacteria [1], which was already recognized as a serious threat to the progress of modern medicine as bacterial infections can become fatal. During COVID-19, the spread of AMR has been exacerbated by infection control lapses, with significantly higher rates of hospital-acquired infections and deaths from multidrug-resistant bacteria in U.S. hospitals [2]. It is estimated that infections from multidrug-resistant bacteria could cause more than 10 million deaths per year worldwide by 2050 [1].

Following the discovery of penicillin in 1928, the Golden Age of antibiotic discovery between the 1940s and the 1960s was led by a systematic survey of microbe-derived antibacterial compounds [3]. During this era, the study discovered numerous antibiotic compounds, such as neomycin and streptomycin produced by soil-dwelling actinomycetes. Most clinically relevant classes of antibiotic compounds were derived from small-molecule natural products, but the excessive use of these compounds resulted in the rapid rise of AMR. Since the 1970s, most antibiotics in clinical trials are derivatives of these antibiotic classes, with a few recent discoveries from bacteria dwelling in the newly-explored environments thanks to the advances in genome mining and pathway analysis [4–6]. According to a recent survey, several dozens of small-molecule antimicrobial candidates are in clinical trials since 2000 - however, only five are first-in-class with a new mechanism of action but none with Gram-negative activity [7]. The majority of the WHO list is gram-negative bacteria (9 out of 12), as they possess an outer membrane that gives resistance to a wide range of antibiotics, including β-lactams, quinolones, and colistins [8]. Particularly, new antibiotics against the carbapenem-resistant gram-negative bacteria (e.g. *Acinetobacter*, *Pseudomonas*, Enterobacteriaceae) are critical, as carbapenems are often used to treat multidrug-resistant infections. For example, up to 7% of Enterobacteriaceae are now resistant to carbapenem due to the rapid spread of extended-spectrum β-lactamase producing strains, causing high morbidity and mortality worldwide [8]. The main barrier for a wide range of antibiotics against gram-negative bacteria is the inability to pass through the outer membrane. For example, hydrophilic antibiotics like β-lactams have to pass through porins, thus any alteration in the outer membrane and its components leads to resistance [8].

Despite the severity of AMR-related issues, too few antibiotics are currently in research and development to counteract the rapid rise in AMR. Between the 5-year period from 2014-2018, only 10 new chemotherapeutic antibiotics have been approved, and none of them targets the critical gram-negative bacteria, such as carbapenem-resistant Enterobacteriaceae (CRE), *Pseudomonas aeruginosa* (CRPA), and *Acinetobacter baumannii* (CRAB) [9]. This collapse in the antibiotic pipeline has been attributed to the absence of a viable antibiotic market, as new antibiotics can only be used sparingly to minimize the risk of further resistance emerging and spreading, despite the huge investments and high risk involved in developing a new drug. We are currently in urgent need of revolutionary next-generation antibiotics that can shift the paradigm of traditional antibiotics, which are mostly broad-spectrum small molecules against which microbes quickly develop resistance.

A recent WHO report on the antibacterial agents in clinical and preclinical development defines traditional antibacterials as small molecules that directly inhibit the growth of (bacteriostatic) or kill bacteria (bactericidal) by targeting essential components for bacterial survival [10]. It also defines non-traditional antibacterials as any other approaches for the treatment and prevention of bacterial infections, or preventing the development or spread of drug resistance. This report presents an analysis of antibacterial agents in preclinical and clinical development worldwide, covering both traditional and non-traditional antibiotics (Figure 1a). As of 2021, there are 46 traditional antibiotics and 34 non-traditional antibiotics in clinical development worldwide. For example, an antibody drug conjugate (ADC) is in clinical development as antimicrobial, which is an engineered human immunoglobulin G1 (IgG1) designed to cleave in phagocytic cells known as a reservoir for *Staphylococcus aureus* infections [11]. Other alternatives to traditional antibiotics are also in development, including phage-based therapy and proteins (Table 2).

The consensus of the AMR experts is that the golden era of antibiotic discovery has passed, as the continuous and systematic study of microbe-derived small-molecule compounds led to no further discovery despite the advances in genomics, bioinformatics, combinatorial chemistry, and high-throughput screening [12]. Since the discovery of penicillin in 1928, the scientific communities possess a much broader and deeper knowledge of microbes in terms of their genome, evolution, ecosystem, and host-parasite interactions. Given the progress in microbial knowledge and technology, the solution to AMR should not be limited to microbe-derived small molecules. The advantages of small-molecule compounds are considerable, as they are easy to manufacture, store, deliver, and administer [3]. However, as these compounds have inherent disadvantages of being non-evolvable, non-specific, and immunogenic, it is essential to develop other types of antibiotics that may not be as convenient but with innovative properties that compensate for the challenges. In this review, we present three innovations that next-generation antibiotics should be differentiated from traditional antibiotics as evolvability, specificity, and non-immunogenicity (Figure 2). Evolvability enables next-generation antibiotics to be updated as bacteria adapt to counteract or evade these antibacterial agents. Specificity allows these antibacterial agents to have minimal off-target effects on human microbiota. Non-immunogenicity reduces negative impact on the human cells and tissues during the antimicrobial treatment. Followingly, we discuss each property in terms of traditional small-molecule antibiotics and non-traditional antimicrobial agents, and present examples of innovation that could overcome the fundamental issues of traditional antibiotics in combating the current AMR crisis.

## 2. Evolvability

2.1. Evolution of antimicrobial resistance in bacteria

Microbes are the most abundant and diverse life forms on Earth, being the most ancient root of life that stretches back 4 billion years ago [13]. It is estimated that only 1% of bacterial and archaeal species have been sequenced and cultured, and the rest of the microbial genomes remain unexplored as Microbial Dark Matter [14]. The evolutionary processes of microbes and viruses are distinctive from those of other higher organisms, as they experience high selective pressures and severe population fluctuations that may be amplified if they have within-host and between-host life cycles [15–17]. Most conventional antimicrobial compounds are derived from bioactive natural molecules, resulting from the interaction of diverse organisms to survive and thrive in nature [18]. Microbes are prolific producers of bioactive natural molecules, particularly soil-dwelling bacteria that make antimicrobial compounds to compete with other microbes or to use as signaling molecules with close relatives or eukaryotic hosts like plants and insects [19]. Thus, antimicrobial resistance is ancient, and the emergence of drug resistance to these antibiotic compounds is intrinsic to the evolutionary processes of complex ecological interactions. Most microbes also undergo pervasive horizontal gene transfer, which facilitates genetic information flow between individuals and populations. This promiscuous exchange of genetic information between microbes is central to the spread of drug resistance, while high mutation rates and short generation times are advantageous for the emergence of drug resistance.

Mutations create new genetic variation, which is fundamental to all evolutionary processes. Mutations that confer resistance to antimicrobials often lower the fitness of a microbe in general, but the high selective pressure of antimicrobial compounds selects for microbes with resistant mutations. These mutations act through diverse strategies, such as modifying the antimicrobial target, regulating the membrane permeability by decreasing antibiotic influx or activating efflux mechanisms, and changing metabolic pathways via regulatory networks [18]. Horizontal gene transfer allows antimicrobial genes to transfer between microbial cells or populations. It is often responsible for the spread of AMR as it allows a rapid introduction of high-fitness-cost antimicrobial genes in a population that is facing a new challenging environment of antibiotics [20]. There are three types of horizontal gene transfer: transformation by which bacteria incorporate foreign genetic materials from the environment, transduction by which bacteria acquire genetic materials through bacteriophages, and conjugation by which DNA is transferred from a donor to a recipient by direct contact. In hospital-acquired infections, conjugation is likely to be the most efficient method of AMR gene transfer due to the high occurrence of microbial cell-to-cell contact, as shown in a recent study within the gastrointestinal tract of humans [18]. Although direct gene transfer between chromosomes has been observed [21], conjugation mostly uses mobile genetic elements like plasmids and transposons as vehicles to disseminate AMR genes. Finally, integrons are site-specific recombination systems that are one of the major drivers of AMR spread. They add new genes and related genes into bacterial chromosomes by recruiting open reading frames in the form of mobile gene cassettes.

The modes of action of most antibiotics can be categorized into five major classes: cell wall, protein synthesis, DNA synthesis, RNA synthesis, and metabolic pathway inhibitors (Table 3) [22]. Resistance to one antibiotic class can result from multiple biochemical pathways, and bacteria are capable of using a combination of resistance mechanisms to escape the effect of an antibiotic. For instance, resistance to fluoroquinolone that blocks DNA synthesis may develop from mutations in genes encoding DNA gyrase and topoisomerase IV, over-expression of efflux pumps, or protection of the protein target sites by another protein (named Qnr) [18]. Due to the difference in the cell envelope, gram-positive bacteria and gram-negative bacteria may differ in the predominant mechanism of resistance. For instance, β-Lactam is a major class of antibiotics that inhibit cell wall synthesis, and gram-positive bacteria mainly modify the penicillin-binding proteins, while gram-negative bacteria produce β-lactamases as their outer membrane can control the access of these antibiotics to the periplasmic space [18]. In overall, the biochemical routes conferring antibiotic resistance can be classified into modifying the antibiotic molecule, preventing access to the target site, changing the target site, and adjusting global cell adaptive processes. Bacteria also possess phenotypic resistance which is non-genetically encoded and non-inheritable resistance to antibiotics through processes such as persistence, biofilms, swarming, and metabolic dormancy [23]. For instance, biofilm is a bacterial population growing in a matrix composed of polysaccharides, proteins, and extracellular DNA, whose structure renders individual cells less susceptible to antibiotics, and persistence occurs when a bacterial subpopulation persists under the presence of antibiotics. For details on the antibiotic resistance mechanisms, the readers are directed to the following review [18].

## 2.2. Phage therapy

Bacteriophages (phages) are viruses that infect and replicate in bacteria, which are the most abundant biological agent on Earth [24]. Lytic or virulent phages infect and kill their bacteria hosts (lytic cycle), whereas lysogenic or temperate phages either integrate into their host's genome (lysogenic cycle) or enter the lytic cycle. As phages are natural killers of bacteria in their lytic cycle, the administration of virulent phages was experimented on early in the 20th century to treat a number of bacterial infections such as cholera, dysentery, bubonic plague, conjunctivitis, and skin infections [25]. The discovery of penicillin in 1929 diminished scientific interest and investment in phage therapy, as a string of cheap and effective antibiotics were introduced to treat bacterial infections. However, phage therapy was steadily developed in places such as Georgia and Poland, which documented extensive and successful cases of phage therapy to treat multiple bacterial infections [26]. Less than a century after the discovery of penicillin, excessive use of antibiotics has resulted in the uncontrolled spread of superbugs, and the lack of new antibiotic discovery renewed therapeutic interest in the potential of phage therapy.

This renewed interest in phage therapy has driven the scientific communities to investigate and standardize various aspects of phage therapy. The minimum regulations for the therapeutic use of phages require strictly lytic phages with antimicrobial activity against the target bacteria and the removal of toxic bacterial debris [27]. Among the standardization, class phage therapy identifies and isolates naturally occurring phages, which are screened for host ranges amid pathogenic bacterial strains, and evaluated *in vitro* or *in vivo* tests. The primary phage of interest is *Caudovirales*, which are the most numerous and diverse phages in the biosphere. They have a linear double-stranded genome of 15 to 500 kb, which make specific contacts to the surface receptors of their bacterial host using the tail or tail fibers, or both. Once the phage genome is injected into the host cell, they typically undergo a lytic cycle which results in replications of hundreds of progeny virions. In the recent clinical setting, phage therapy was focused on the clinical product development against bacterial pathogens such as *Staphylococcus aureus*, *Pseudomonas aeruginosa*, and *Clostridium difficile*, which are difficult to treat with conventional antibiotic therapy. For instance, Phagoburn was the world-first phage therapy clinical trial using phage cocktails for the treatment of *Escherichia coli* and *Pseudomonas aeruginosa* burn wound infections, which achieved significant advancements in the regulatory framework of phage therapy [28]. Several companies have already commercialized phage products for controlling food-borne pathogens such as *Escherichia coli O157:H7* and *Listeria monocytogenes*, thanks to the genetic homogeneity of these bacteria and the lower regulatory barriers for food production and processing [29].

Alternative antibiotics still face significant challenges; phage therapy has safety concerns of self-replicating bacteriophages in patients [30] and bacteriophage-derived agents have delivery issues to different organs given the harsh *in vivo* environments (e.g. low pH, cell barriers, proteases) [31]. Despite these challenges, the natural antimicrobial activities of bacteriophages are gaining attention as viable alternatives; phage therapy is actively being tested in clinical trials (Table 2) and is already in use to treat bacterial infections in some countries like Georgia [32]. Several phage-encoded endolysins which lyse the bacterial peptidoglycan layer are in clinical development against Gram-positive bacteria (Table 2) [31]. However, no phage-based antimicrobial agents have been approved yet, due to regulatory and logistical hurdles [33]. Currently, no bacteriophage-based therapeutics have passed FDA approval for clinical use, except in emergency or experimental cases [34].

## 2.3. Evolvability of a bacteriophage-based therapy

Unlike chemical-based traditional antibiotics, bacteriophages are biological entities that are self-replicating and evolving under changing environments. This characteristic is both an advantage and disadvantage to controlling bacterial populations. This paradoxical relation also stands in natural environments where diminishing bacterial populations due to highly successful infections of lytic phages will eventually diminish the chance of their own replication too. During the host-parasite interaction, bacteria can develop resistance to phage infections, equivalent to the case of antibiotics. The difference is, however, phages also evolve to counteract the defense systems of bacteria, whose evolution can be directed and accelerated through genetic engineering to outpace the bacterial resistance and even enhance their replication and lytic activities. Previously, phages were engineered to add or improve function as natural predators of bacteria, such as an engineered enzymatic bacteriophage incorporated with a gene that degrades a polysaccharide adhesin in biofilm formation [35].

Bacteria have various defense mechanisms against these phages, such as restriction-modification systems that protect host DNA with modification and destroy foreign DNA with restriction enzymes [36], and CRISPR-Cas systems that specifically degrade previously encountered foreign genetic elements through RNA templates [37]. However, phages also have several arsenals to counteract these bacterial defense systems. For instance, recent studies revealed that phages have small proteins that have anti-CRISPR activities by inhibiting CRISPR-Cas systems via direct interference [38,39] or enzymatic activity [40,41]. Bacterial populations may develop a collective strategy to mitigate phage infection, such as a newly discovered system named CBASS (cyclic oligonucleotide-based anti-phage signaling system) that uses small signaling molecules to activate cell death released upon phage infection [42]. Such diverse bacterial defense strategies may result in unexpected results such as the depletion of phage replications during phage therapy. Furthermore, bacteria may adapt other phenotypic and genotypic changes like decreased phage absorption due to the intense selective pressure imposed by phages [43], which may require other treatment strategies such as the use of phage cocktails and phage engineering [44].

Phages have been known to be highly specific for their hosts, which enables the targeting of pathogenic bacteria at the strain level without disturbing microbiomes in the body [45]. The inevitable off-target effects from conventional antibiotic therapy are known to cause severe disruptions in the microbiomes of the human body (see below for details). However, there is recent evidence that phages can also jump hosts, and this adaptation of phage-host specificity may lead to unexpected loss or gain in specificity [46]. The host receptor should be identified for any phage proposed for therapeutic use, to minimize off-target events and also to assemble combinations of phages that are less likely to generate resistant hosts that are defective in a single receptor. Furthermore, the use of lysogenic phages in phage therapy should be prohibited, as they can carry antimicrobial or virulence genes that alter the pathogenic potential of their hosts [47]. Thus, the evolvability of bacteriophage-based therapy is advantageous in counteracting highly adaptive bacteria. On the other hand, it brings unpredictability and instability to the antimicrobial treatment, and requires constant monitoring and evaluation to minimize uncertainty.

# 3. Specificity

## 3.1. Broad-spectrum and narrow-spectrum small-molecule antibiotics

Small-molecule antibiotics have variable ranges of microorganisms they can inhibit. Based on the spectrum of antimicrobial activity, they can be divided into broad-spectrum antibiotics that can target a wide range of bacteria or narrow-spectrum antibiotics that can target limited species of bacteria. Extended-spectrum antibiotics can target gram-positive bacteria but only some gram-negative bacteria (Table 3). Generally, broad-spectrum antibiotics have higher chances of developing antimicrobial resistance, as the selective pressure for resistance is applied on both pathogenic and non-pathogenic bacteria. During this process, non-pathogenic commensal bacteria in microbiomes become a persistent reservoir for antimicrobial resistance genes that can be transferred to pathogenic bacteria [48]. Broad-spectrum antibiotics not only promote the emergence of multidrug-resistant bacteria and cause dysbiosis in the microbiome from off-target effects, but they also have more side effects such as diarrhea or rash [49]. Thus, antibiotic stewardship generally recommends identifying the specific pathogen and facilitating the use of narrow-spectrum antibiotics over broad-spectrum antibiotics whenever possible, although broad-spectrum antibiotics tend to have more clinical indications.

The development of narrow-spectrum antimicrobial agents that are genus or species-specific is one of the strategies to tackle antimicrobial resistance. Narrow-spectrum antibiotics are less likely to induce antimicrobial resistance and disrupt the human microbiome [48]. This is an important advantage as the effects of antibiotic exposure as short as seven-day have been shown to alter the gut microbiota over two years post-treatment [50]. A recent study demonstrates that repeated use of antibiotics may permanently change the size and composition of gut microbiomes [51]. As microbiomes play vital roles in human physiology, such as protection from pathogens and metabolite production, microbiome dysbiosis leads to disruptions to human health (see below for details). Particularly, it has been observed that exposure to broad-spectrum antibiotics during early childhood disrupts the diversity and stability of the infant microbiota, which can also be disruptive to the development of the infant immune system [52].

Although broad-spectrum antibiotics are essential for life-threatening infections like sepsis or pneumonia, better identification of the causative pathogen allows a switch to narrow-spectrum antibiotics that can reduce both the antimicrobial resistance and the microbiome disruption in non-life-threatening infections like urinary tract infections and abscesses.

## 3.2. Specificity of CRISPR-based antimicrobials

CRISPR-Cas systems are microbial immune systems first discovered in bacteria, which consist of Clustered Regularly Interspaced Short Palindromic Repeats (CRISPR) arrays and CRISPR-associated system (Cas) proteins [37,53]. CRISPR arrays are a remarkable component of CRISPR-Cas systems that enables RNA-mediated adaptive immunity by encoding genetic information about previous invaders such as phages or plasmids. This genetic component makes CRISPR-Cas systems programmable and specific by altering the target information, and they have been successfully adapted as genome-editing tools thanks to this characteristic [54,55]. There are two main classes and several types of CRISPR-Cas systems depending on the architecture of Cas proteins, and the diversity of these prokaryotic immune systems has been expanding as more uncultured microbes from diverse environments are being discovered and sequenced [56].

Among these, a number of CRISPR-Cas systems have been recently investigated as alternative antibiotics by reprogramming them to target bacterial DNA/RNA [57,58]. In a landmark study, the prominent genome-editing tool, CRISPR-Cas9 systems, was repurposed to target multidrug-resistant bacteria [59]. They used bacteriophages and bacterial plasmids to deliver CRISPR-Cas systems encoding virulence and antimicrobial resistance templates in carbapenem-resistant Enterobacteriaceae, which significantly increased the survival rate of the worm infection model organism. In another pioneering study, the CRISPR-Cas9 systems delivered by a plasmid packaged in phage capsids called phagemids were reported to selectively kill virulent strains of *Staphylococcus aureus* [60]. These sequence-specific antimicrobials were validated also in a murine skin infection model. In another study, another type of DNA-modifying CRISPR-Cas system was used to target antibiotic-resistant bacteria by destroying their antibiotic-resistance-conferring plasmids with temperate and lytic phages as delivery vectors [61].

More recently, the potential of type VI CRISPR-Cas systems as antimicrobial tools is gaining attention because these proteins cleave targeted transcripts of invading RNA viruses. Triggered by the target RNA cleavage, type VI CRISPR-Cas systems carry out non-specific RNase activity, resulting in cleaving transcripts of the bacterial genome itself. This activity eventually leads to the dormancy of the bacterial host cell, which disables invading phages from multiplying further into other bacterial cells thus diminishing the phage population. Some recent studies took advantage of this outcome to trigger bacterial cells to enter cell arrest when CRISPR-Cas systems detect the expression of antimicrobial resistance genes [62]. This strategy has advantages over other DNA-modifying CRISPR-Cas systems as there is no need to consider the potential interference of extensive DNA repair systems in bacteria [63].

CRISPR-Cas systems found in prokaryotes are initially applied as genome-editing tools, and their specificity and programmability are also highly attractive traits as alternative antibiotics that traditional antibiotics do not possess. If we could successfully repurpose CRISPR-Cas systems as antibiotics, they can be programmed to be specific to pathogenic bacteria at the strain level instead of disturbing the whole microbiota in the human body. Currently, no CRISPR-Cas-based therapeutics have passed FDA approval for clinical use, but there is one clinical development of non-traditional antibiotics involving CRISPR-Cas3 enhanced phages (Table 2). Since CRISPR-Cas antimicrobials are to be used against bacteria that have their own CRISPR-Cas systems, their functioning within bacteria should be investigated further to prevent unexpected events from indigenous CRISPR-Cas and genomic systems.

## 3.3. Diagnostic tests for pathogen identification

One of the main challenges to the utility of next-generation antibiotics with high specificity is the requirement for rapid, accurate, and sensitive diagnosis of bacterial pathogens. To achieve specific targeting of pathogens causing infections, rapid pathogen identification with high accuracy and sensitivity is vital. Due to the availability of broad-spectrum antibiotics, most bacterial infections have been treated without the need for diagnostic tests. Such empirical antibiotic therapy exacerbated the emergence and spread of antimicrobial resistance, which necessitates a shift towards directed antibiotic therapy along with the progress of diagnostic clinical microbiology.

Culture-based diagnosis in clinical microbiology is dependent on the growth of bacteria and has largely been unchanged for 100 years. Culture-based diagnostic processes of bacterial infections take several days, from initial cultures (~24 hours) to pathogen identification and antimicrobial susceptibility testing (~24 hours) [64]. This diagnostic method is also prone to false-negative results, particularly if samples are obtained during antimicrobial therapy. Other techniques such as Gram-staining microscopy and ELISA for detecting bacterial antigens or antibodies are less time-consuming, but they cannot determine antimicrobial susceptibility [65].

In the recent clinical diagnostic setting, the introduction of nucleic acid-based amplification technologies (NAATs) and MALDI-TOF mass spectrometry fingerprinting have modernized pathogen identification [64]. NAAT-based approaches include polymerase chain reaction (PCR) and next-generation sequencing (NGS), which accelerates pathogen detection to within 3-6 hours [64]. PCR-based techniques have higher sensitivity than culture-based approaches, as demonstrated in cases when antimicrobial treatment is ongoing or only small sample volumes are available (e.g. bloodstream infection) [66]. However, PCR-based techniques can lead to false positives due to the presence of genetic materials after the pathogen has been neutralized, and false negatives due to the emergence of mutations or loss of the gene during antibiotic treatment. NGS has similar limitations, but it exhibits increased accuracy with the potential to detect antimicrobial resistance genes and virulence markers. MALDI-TOF

MS fingerprinting uses direct colony testing on the MALDI plate to compare the generated spectrum against a reference spectrum for bacterial pathogen identification [67]. This method is rapid, accurate, and inexpensive, but only clinical samples with high numbers of bacteria, such as urine and cerebrospinal fluid, allow direct testing, and organisms with similar spectral profiles, such as *E. coli* and *Shigella* species cannot be differentiated accurately [67]. Other approaches include fluorescent *in situ* hybridization (FISH) and electrochemical biosensor assays by species-specific probes for the bacterial ribosomal RNA target, and rapid antigen testing by a visible readout upon antibody-antigen binding [48].

# 4. Non-immunogenicity

## 4.1. Effects of antibiotics on the immune system

According to the Centers for Disease Control and Prevention (CDC), the most common side effects of antibiotics involve the digestive system and the immune system. Due to the detrimental effect on microbiota homeostasis, antibiotics can cause nausea, diarrhea, and indigestion. The negative effects of antibiotics also include allergic reactions such as rash, coughing, wheezing, and breathing difficulties. In rare cases, antibiotics can cause a medical emergency such as anaphylaxis, which is a severe and life-threatening allergic reaction. Most emergency department visits related to antibiotic side effects are due to severe allergic reactions.

Infants are vulnerable to bacterial infections, especially when born preterm and/or underweight, and are often subjected to prophylactic or therapeutic antibiotic treatments [68]. It is estimated that around 40% of pregnant mothers and newborns receive antibiotics globally [69]. In fact, empiric antibiotic treatment is a common practice during pregnancy and birth, which leads to the inappropriate use of antibiotics, particularly in developing countries. In infants, the use of antibiotics has been found to cause more long-lasting negative effects on the immune system, which is not fully established and functional. Antibiotic therapy during infancy is linked to a higher risk of infections later in life, as shown in the studies which found associations between prolonged exposure to antibiotics with an increased susceptibility to diarrhea and respiratory tract infections [68].

Using the animal models, some studies demonstrated that antibiotic exposures during infancy negatively impact innate immune cells, such as natural killer (NK) cells, dendritic cells (DCs), and innate lymphoid cells. For instance, the infant mice born from the antibiotic-treated mothers after being infected with the vaccinia virus had a reduced number of splenic DCs, which are the most potent antigen-presenting cells, compared with the control mice [70]. Similarly, NK cells of the antibiotic-exposed mouse infants exhibited remarkable reductions in terms of frequency and phenotypic expression following the vaccinia virus infection. In terms of adaptive immunity, the mouse infants exposed to antibiotics in early life had their antibody-mediated responses impaired to the majority of vaccines, such as protection against tuberculosis, meningitis, and pneumococcal disease, compared to the control mouse infants [71]. Thus, these studies indicate that antibiotic exposure during early life could cause long-lasting impairments both in innate immunity and adaptive immunity.

## 4.2. Human microbiome

A microbiome is a collection of cells, genes, and metabolites from the microbiota comprising bacteria, viruses, and eukaryotes within the human body. The high-throughput technological advances in sequencing and data processing have allowed the scientific community to establish a baseline of healthy microbiome compositions, to which microbiome compositions from patients with various diseases can be compared. A healthy microbiome profile is generalizable across human populations consisting of the commensal and beneficial microbiota [72]. The human microbiome is tightly involved in human health, particularly the human gut is inhabited by trillions of microbes influencing host physiology and susceptibility to diseases, including malnutrition [73], obesity [74], inflammatory bowel disease [75], neurological disorders [76] and even cancer [77]. In addition to the gut microbiome, the complex oral microbiome also plays a key role in maintaining both oral health and systemic health, and its dysbiosis has been linked to a vast array of health issues including respiratory, cardiovascular, and cerebrovascular diseases [78].

      The use of antibiotics causes dysbiosis, which is a disruption to the microbiome from an imbalance in microbiota, activities, or distributions [45], particularly during infancy and early childhood. For instance, antibiotic drugs decrease the overall diversity and increase the colonization of drug-resistant pathogens of gut microbiota in infants [79]. Several studies have linked the repeated use of antibiotics such as penicillins, macrolides, quinolones, and cephalosporins in early childhood to long-term health

issues such as an increased risk of developing obesity [80] and type 2 diabetes [81]. Even in adulthood, antibiotic use has been shown to transiently or permanently affect the diversity and health of human microbiota by depleting several commensal and beneficial taxa such as lactobacilli and bifidobacteria [45]. Furthermore, antibiotics select for resistance in the gut microbiota by stimulating the expression of antibiotic resistance, stress response, and virulent phage genes [82]. There is also evidence that antibiotics can cause immunological disorders by negatively impacting the interaction between the microbiome and immune system [83] and perturbing the host proteome [84].

## 4.3. Non-immunogenicity of microbiome-derived antibiotics

Microbes inhabiting the same environmental niches within human microbiomes develop various strategies to gain advantages over other microbes. The human microbiota is known to produce a diverse spectrum of metabolites specific for interacting within the human microbiota and human hosts, such as lipids, oligosaccharides, amino acids, non-ribosomal peptides, and ribosomal peptides [85]. These metabolites serve a variety of purposes, including antimicrobial, cytotoxic, immunomodulatory, and antioxidant functions. The human microbiome has revealed several natural products with antimicrobial properties across the bacterial phyla, such as the vaginal isolate lactocillin and the nasal isolate lugdunin [86,87].

     The human gut is a particularly dense environment where trillions of bacteria, archaea, eukarya, and viruses coexist and coevolve, and this competition has led to various strategies to outcompete others, including the development of specialized antimicrobials. In the gut microbiome, some bacteria use direct antagonistic strategies against their neighbors, such as removing essential substrates, reducing oxidation-reduction potential, and accumulating D-amino acid [85]. More indirect strategies involve the production of metabolic compounds that limit the growth of surrounding bacteria. For example, some bacteria produce hydrogen peroxide, which is a non-specific regulatory agent with antimicrobial activities through

oxidizing effects on bacterial molecular structures [88]. However, due to the non-specific activity and associated side effects like the acidification of the environment, most of these bacterial compounds with antimicrobial activities are unsuitable for clinical applications.

Bacteria also produce antimicrobial peptides consisting of 10-50 amino acids that are target-specific. The ability of these peptides to neutralize bacteria depends on their affinities to bacterial membranes and cell walls [85]. The first category of microbiome-derived antimicrobials is non-ribosomal peptides (NRP) which are secondary metabolite peptides synthesized by multifunctional peptide synthetases. Several microbe-derived antibacterial compounds, including penicillin, vancomycin, and polymyxin, are considered nonribosomal peptides. However, most activities of microbiome-derived NRPs are known to be cytotoxic and only a few NRPs have been characterized from the human microbiota [89].

The second category of microbiome-derived antimicrobials is ribosomally synthesized peptides that were first discovered in 1925 and referred to as bacteriocins [85]. Generally, bacteriocins produced by Gram-positive bacteria work better against Gram-positive pathogens and Gram-negative bacteriocins against Gram-negative pathogens. Bacteriocins are heterogeneous in primary structure, molecular weight, mode of action, and heat stability, and the most current classification is based on their structure [90]. Bacteriocins have low toxicity in human cells with broad-spectrum or narrow-spectrum antimicrobial activities against bacterial cells. Another advantage of bacteriocins as antimicrobial compounds is that bacteria cannot easily develop resistance against them as these pathogens have to alter their membrane or receptor compositions. Recently, antimicrobial peptides from the rumen microbiome exhibited therapeutic potential against seven clinical strains of *Pseudomonas aeruginosa* with minimal cytotoxicity against human lung cells [91]. These antimicrobial peptides increased catalytic activities at the target bacterial cell membrane and promoted the β-oxidation of fatty acids. This study illustrates the therapeutic potential of microbiome-derived non-ribosomal peptides against bacterial infections.

Recent evidence reveals that diverse and numerous bacteriophages coexist in the human body without causing immunogenic reactions [92,93]. Some bacteriophages produce lytic enzymes that can kill bacteria, and the use of the phages derived from the human gut has been proposed as a novel therapeutic to modulate gut composition [94,95]. Phages are inherently harmless to eukaryotic cells, but they can cause immunological reactions due to the bacterial lysates and endotoxins resulting from the phage lytic cycles. Microbiome-derived bacteriophages are largely unexplored and the uncharted repertoire of bacteriophages is a rich resource for genome mining of next-generation antibiotics [96]. Two phage-derived peptides are potential antibacterial therapeutics: lysins and tailocins. Lysins are muralytic enzymes that are used both at the early stage of infection to penetrate the DNA through the host cell envelope and at the lysis stage of infection to release the progeny virions [97]. These enzymes are effective against Gram-positive bacteria with high genus-level specificity. Tailocins are phage tail-like bacteriocins that cause lethal damage to the host cell envelope upon absorption into a bacterial surface receptor [98]. Tailocins are inherently devoid of genetic materials and can be engineered to target heterologous hosts that can be administered at a defined dose. Currently, the microbiome-modulating category has the highest number of non-traditional antimicrobial agents in clinical development, with one agent in the most advanced stage of new drug application (Figure 1b).

# 5. Conclusion

Antimicrobial resistance is an imminent threat to global public health, but the interest and investment in the issue are not proportional to the scale and severity of the crisis. We are in urgent need of new antibiotics, as the antibiotics have lagged behind the rapid adaptation of bacteria. Microbes will always evolve, but we have been dependent mostly on microbe-derived small molecules to counteract the ingenious strategies of microbes to evade these antimicrobial compounds. As our scientific knowledge of microbes and their complex communities expands, it is essential to diversify strategies for combating pathogenic multidrug-resistant bacteria that can cause fatal infections. In this review, we presented three innovations that next-generation antibiotics can focus on as evolvability, specificity, and non-immunogenicity. We used some examples of alternative antimicrobial agents currently in research or clinical development to expand on the properties of each innovation, including bacteriophage-based therapy, CRISPR-Cas-based antimicrobials, and microbiome-derived antimicrobial agents. It should be noted that there are other innovative agents such as antibodies and nanoparticles that are being developed as alternative antimicrobials (Table 2).

The research and development of next-generation antibiotics are affected by profitability challenges, as the market size of a drug is proportional to the prevalence of the disease. However, the evolvability and specificity of next-generation antibiotics may compensate for such reduced revenue due to the lower prevalence by lengthening the viability of antimicrobials with reduced rates of resistance. If the three innovations of next-generation antibiotics can overcome the challenges of traditional antibiotics and are more effective in reducing mortality, morbidity, and length of hospitalization, the case for a higher price may also be made in high-income countries.

To prepare for the imminent post-antibiotic era, a shift in medical culture and education from empirical antibiotic therapy to directed antibiotic therapy is necessary. Furthermore, a shift in the pharmaceutical industry to invest in innovative next-generation antibiotics rather than broad-spectrum small-molecule antibiotics is essential. Regulatory changes by governmental agencies such as the Food and Drug Administration (FDA) and the European Medicines Agency (EMA) to support and monitor research and development of next-generation antibiotics are required to encourage a new era of antibiotics that are specific, evolvable, and non-immunogenic. Next-generation antibiotics still have technological limitations and regulatory hurdles to overcome. However, they represent another scientific asset that will progress modern medicine by shifting the paradigm of antibiotics from exclusively chemotherapeutic small molecules to a diversified range of tools and agents to combat antimicrobial resistance that humanity will continue to face. In closing, further research to explore novel microbes and microbial communities is essential to get inspiration for novel antimicrobial strategies to be repurposed as next-generation antibiotics.

## Consent for publication

Not applicable.

## Funding

The research and development activities described in this study were funded by Ghent University Global Campus (GUGC), Incheon, Korea.

## Declaration of Competing interests

None

## Acknowledgements (optional)

The author thanks the members of the Center for Biotech Data Science at GUGC for constant encouragement, support, and motivation.

**Figure 1: Antibiotics in clinical development according to the WHO analysis (published in 2022). a**, Traditional antibiotics versus next-generation antibiotics. **b**, Next-generation antibiotics by antibacterial class and development phase.

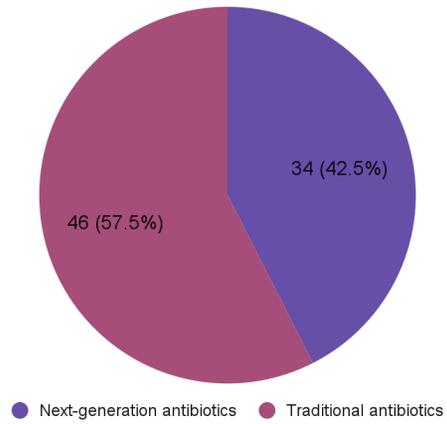

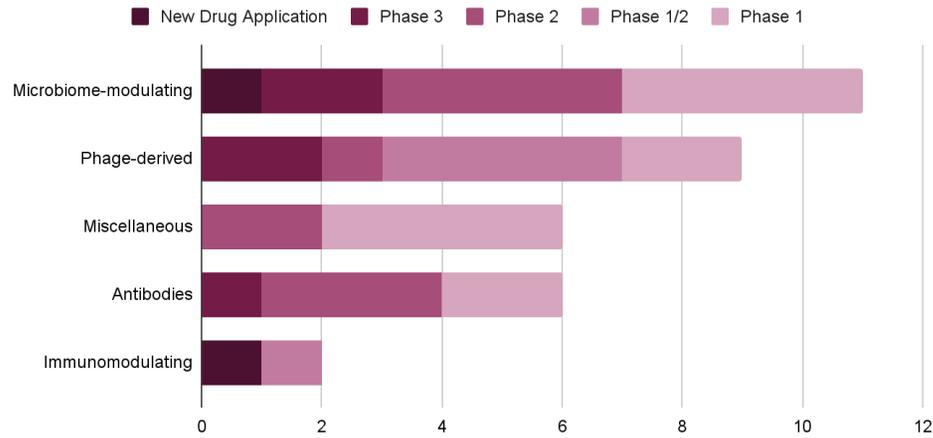

**Figure 2: Three innovations of next-generation antibiotics.**

|  | Evolvability | Specificity | Non-immunogenicity |
|---|---|---|---|
| **Traditional antibiotics** | become obsolete as bacteria continuously evolve to develop AMR | are generally broad-spectrum, causing microbiota dysbiosis | have one of the main side effects as allergic reactions |
| **Next-generation antibiotics** | are adaptable to counteract as bacteria develop AMR e.g. phage therapy | enable species-specific or strain-specific targeting of bacteria e.g. CRISPR-based antimicrobials | are compatible with the human body and immune system e.g. microbiome-derived compounds |

Table 1: World Health Organization (WHO) priority pathogens for R&D of new antibiotics (released in 2017) and Centers for Disease Control and Prevention (CDC) antibiotic resistance threats in the United States (released in 2019).

| WHO | Pathogen | Type | CDC | | Type |
|---|---|---|---|---|---|
| **Priority 1: CRITICAL** | *Acinetobacter baumannii*, carbapenem-resistant | Gram-negative bacteria | **Urgent Threats** | Carbapenem-resistant *Acinetobacter* | Gram-negative bacteria |
| | *Pseudomonas aeruginosa*, carbapenem-resistant | Gram-negative bacteria | | *Candida auris* | Fungus |
| | Enterobacteriaceae, carbapenem-resistant, 3rd gen. cephalosporin-resistant | Gram-negative bacteria | | *Clostridioides difficile* | Gram-positive bacteria |
| **Priority 2: HIGH** | *Enterococcus faecium*, vancomycin-resistant | Gram-positive bacteria | | Carbapenem-resistant *Enterobacterales* | Gram-negative bacteria |
| | *Staphylococcus aureus*, methicillin-resistant, vancomycin-resistant | Gram-positive bacteria | | Drug-resistant *Neisseria gonorrhoeae* | Gram-negative bacteria |
| | *Helicobacter pylori*, clarithromycin-resistant | Gram-negative bacteria | **Serious Threats** | Drug-resistant *Campylobacter* | Gram-negative bacteria |
| | *Campylobacter species*, fluoroquinolone-resistant | Gram-negative bacteria | | Drug-resistant *Candida* | Yeast |
| | *Salmonellae species*, fluoroquinolone-resistant | Gram-negative bacteria | | ESBL-producing *Enterobacterales* | Gram-negative bacteria |
| | *Neisseria gonorrhoeae*, 3rd gen. cephalosporin-resistant, fluoroquinolone-resistant | Gram-negative bacteria | | Vancomycin-resistant *Enterococci* (VRE) | Gram-positive bacteria |
| **Priority 3: MEDIUM** | *Streptococcus pneumoniae*, penicillin-non-susceptible | Gram-positive bacteria | | Multidrug-resistant *Pseudomonas aeruginosa* | Gram-negative bacteria |
| | *Haemophilus influenzae*, ampicillin-resistant | Gram-negative bacteria | | Drug-resistant nontyphoidal *Salmonella* | Gram-negative bacteria |
| | *Shigella species.*, fluoroquinolone-resistant | Gram-negative bacteria | | Drug-resistant *Salmonella serotype Typhi* | Gram-negative bacteria |
| | | | | Drug-resistant *Shigella* | Gram-negative bacteria |
| | | | | Methicillin-resistant *Staphylococcus aureus* (MRSA) | Gram-positive bacteria |
| | | | | Drug-resistant *Streptococcus pneumoniae* | Gram-positive bacteria |
| | | | | Drug-resistant *Tuberculosis* | Gram-positive bacteria |

| | | |
|---|---|---|
| **Concerning Threats** | Erythromycin-Resistant Group A *Streptococcus* | Gram-positive bacteria |
| | Clindamycin-resistant Group B *Streptococcus* | Gram-positive bacteria |
| **Watch List** | Azole-resistant *Aspergillus fumigatus* | Fungus |
| | Drug-resistant *Mycoplasma genitalium* | Gram-negative bacteria |
| | Drug-resistant *Bordetella pertussis* | Gram-negative bacteria |

**Table 2: Next-generation antibiotics in clinical development according to the WHO analysis (published in 2022).**

| Class | Name (synonym) | Phase | Antibacterial class | Route of administration | Expected activity against priority pathogens |
|---|---|---|---|---|---|
| Phage-derived | Exebacase (CF-301) | 3 | Phage endolysin | intravenous | *S. aureus* |
| | Bacteriophage cocktail | 3 | Phage | inhalation | Gram-positive and Gram-negative |
| | LSVT-1701 (N-Rephasin SAL200, tonabacase) | 2a/1 | Phage endolysin | intravenous | *S. aureus* |
| | Phage | 1/2 | Phage | intravenous | *E. coli* |
| | AP-PA02 | 1/2 | Phage | inhalation | *P. aeruginosa* |
| | YPT-01 | 1/2 | Phage | inhalation | *P. aeruginosa* |
| | BX004-A | 1/2 | Phage | inhalation | *P. aeruginosa* |
| | LBP-EC01 | 1b | CRISPR-Cas3 enhanced phage | intravenous | *E. coli* |
| | LMN-201 | 1b | Phage endolysin and three toxin-binding proteins (5D, E3 and 7F) | oral | *C. difficile* |
| Microbiome-modulating | BB128 | Marketing Authorization Application | Live biotherapeutic product | colonoscopy | *C. difficile* |
| | SER-109 | 3 | Live biotherapeutic product | oral | *C. difficile* |
| | RBX2660 | 3 | Live biotherapeutic product | enema | *C. difficile* |
| | SYN-004 (ribaxamase) | 2b | Antibiotic inactivator | oral | *C. difficile* |
| | VE303 | 2 | Live biotherapeutic product | oral | *C. difficile* |
| | CP101 | 2 | Live biotherapeutic product | oral | *C. difficile* |
| | DAV132 | 2 | Antibiotic inactivator and protective colon-targeted adsorbent | oral | *C. difficile* |
| | MET-2 | 1 | Live biotherapeutic product | oral | *C. difficile* |
| | RBX7455 | 1 | Live biotherapeutic product | oral | *C. difficile* |
| | ART24 | 1 | Live biotherapeutic product | oral | *C. difficile* |
| | SVT-1C469 | 1 | Live biotherapeutic product | oral | *H. pylori* |
| Immunomodulating | Reltecimod (AB103) | New Drug Application | Synthetic peptide antagonist of both superantigen exotoxins and the CD28 | intravenous | *S. aureus* |

|  | | | T-cell receptor | | |
|---|---|---|---|---|---|
|  | Rhu-pGSN (rhu-plasma gelsolin) | 1b/2a | Recombinant human plasma gelsolin protein | intravenous | Non-specific Gram-positive and Gram-negative |
| Antibodies | Tosatoxumab (AR-301) | 3 | Anti-*S. aureus* IgG1 antibody | intravenous | *S. aureus* |
|  | LMN-101 | 2 | mAb-like recombinant protein | oral | *E. coli, C. jejuni* |
|  | AR-302 (MEDI4893, suvratoxumab) | 2 | Anti-*S. aureus* IgG mAb | intravenous | *S. aureus* |
|  | IM-01 | 2 | Chicken egg-derived anti-*C. difficile* polyclonal antibody | oral | *C. difficile* |
|  | TRL1068 | 1 | mAB | intravenous | Gram-positive and Gram-negative biofilms |
|  | 9MW1411 | 1 | mAb (α-toxin) i | intravenous | *S. aureus* |
| Miscellaneous | OligoG (CF-5/20) | 2b | Alginate oligosaccharide (G-block) fragment | inhalation | *P. aeruginosa* |
|  | Ftortiazinon (fluorothyazinone) + cefepime | 2 | Thyazinone (type III secretion system inhibitor) + cephalosporin | oral | *P. aeruginosa* |
|  | CAL02 | 1 | Broad-spectrum anti-toxin liposomal agent and nanoparticle | intravenous | *S. pneumoniae* |
|  | BVL-GSK098 | 1 | Amido piperidine (inactivation of TetR-like repressor EthR2) | oral | *M. tuberculosis* |
|  | GSK3882347 | 1 | Undisclosed (FimH antagonist) | oral | *E. coli* |
|  | ALS-4 | 1 | Anti-virulence (staphyloxanthin biosynthesis inhibition) | oral | *S. aureus* |

**Table 3: Mechanism of action and sensitivity against gram-negative bacteria of each antibiotic group.**

| Mechanism of Action | Antibiotic Group | Examples | Gram(-) coverage |
|---|---|---|---|
| Inhibit Cell Wall Synthesis | β-Lactams | Penicillins, Cephalosporins, Monobactams, Carbapenems | Some |
| | Glycopeptides | Vancomycin, Teicoplanin | No |
| Depolarize Cell Membrane | Lipopeptides | Daptomycin, Surfactin | No |
| Inhibit Protein Synthesis | Bind to 30S Ribosomal Subunit | Aminoglycosides, Tetracyclines | Yes |
| | Bind to 50S Ribosomal Subunit | Chloramphenicol, Lincosamides, Macrolides, Oxazolidinones, Streptogramins | Some |
| | Fusidic Acid | | No |
| Inhibit DNA Synthesis | Quinolones | Fluoroquinolones | Yes |
| | Metronidazole | Metronidazole, Tindazole | Yes |
| | Nitrofurantoin | Nitrofurantoin, Furazolidone | No |
| Inhibit RNA Synthesis | Ansamycins | Geldanamycin, Rifamycin, Naphthomycin | Yes |
| Inhibit Metabolic Pathways | Sulfonamides | Prontosil, Sulfanilamide, Sulfisoxazole | Yes |
| | Trimethoprim | Sulfasalazine, Sulfadiazine | Yes |